\documentclass[letter,twocolumn]{jpsj2} %% two-column layout
\title{Substitution Effect by Deuterated Donors on Superconductivity in
$\kappa$-(BEDT-TTF)$_2$Cu[N(CN)$_2$]Br}

\author{Naoki \textsc{YONEYAMA}$^{1}$\thanks{E-mail address: 
yone@imr.tohoku.ac.jp}, Takahiko \textsc{SASAKI}$^{1}$
and Norio \textsc{KOBAYASHI}$^{1,2}$} 

\inst{$^{1}$Institute for Materials Research, Tohoku University, Sendai
980-8577, Japan \\
$^{2}$Center for Low Temperature Science, Tohoku University, Sendai 
980-8577, Japan}

\abst{We investigate the superconductivity in the deuterated BEDT-TTF 
molecular substitution system 
$\kappa$-[(h8-BEDT-TTF)$_{1-x}$(d8-BEDT-TTF)$_x$]$_2$Cu[N(CN)$_2$]Br, 
where h8 and d8 denote fully hydrogenated and deuterated molecules, 
respectively. 
Systematic and wide range ($x$ = 0 -- 1) substitution can control 
chemical pressure finely near the Mott boundary, which results in the 
modification of the superconductivity.
After cooling slowly, the increase of $T_{\textrm{c}}$ observed up to $x 
\sim$ 0.1 is evidently caused by the chemical pressure effect. 
Neither reduction of $T_{\textrm{c}}$ nor suppression of 
superconducting volume fraction is found below $x \sim$ 0.5.
This demonstrates that the effect of disorder by substitution is 
negligible in the present system.
With further increase of $x$, both $T_{\textrm{c}}$ and superconducting 
volume fraction start to decrease toward the values in $x$ = 1.
}

\kword{organic superconductor, deuterium substitution, chemical pressure, Mott 
transition, cooling rate}

\begin{document}
\maketitle

The $\kappa$-type family of organic charge transfer salts, 
$\kappa$-(BEDT-TTF)$_2X$, has a two-dimensional crystal structure
consisting of alternating layers of conducting donors and insulating 
anions,\cite{Ishiguro} where BEDT-TTF (ET) is 
bis(ethylenedithio)tetrathiafulvalene.
The strongly correlated electron interaction plays an important role in 
the electronic states of this system, which is quite sensitive to 
hydrostatic pressure, anion, and deuterium substitution.
In the salt with $X$ = Cu[N(CN)$_2$]Cl, whose ground state is an 
antiferromagnetic insulating (AFI) state at ambient pressure, 
a superconducting (SC) state appears by pressure of a few hundred bar.\cite{235,327,986,971}
This insulator-to-metal transition has been explained as a Mott transition under 
the control of the conduction band width.\cite{346}
The Mott transition can also be induced by substituting Br for Cl;
the ground state of the salt with $X$ = Cu[N(CN)$_2$]Br (hereafter 
abbreviated as the Cu[N(CN)$_2$]Br salt) is a SC one at ambient pressure.
This change of the ground state by anion substitution
is caused by the larger anion size of Br than that of Cl (chemical pressure).
However, for the appearance of the bulk SC state in the Cu[N(CN)$_2$]Br salt, the 
conventional hydrogenated ET (h8-ET) donor is needed. 
It is because deuterium substitution in the h8-ET, forming the deuterated ET (d8-ET), 
results in an AFI ground state coexisting with a minor SC phase.\cite{144,749}
This metal-to-insulator transition by deuterium substitution is 
attributed to (negative) chemical pressure, originated from 
the slight difference between the C--H and C--D bond lengths.
In addition, deuterium substitution increases $T_{\textrm{c}}$ by 0.3 K 
in the SC salt with $X$ = Cu(NCS)$_2$ (the 
Cu(NCS)$_2$ salt).\cite{926}
Even in other superconductors, i.e.,  $\beta$''-(ET)$_2$SF$_5$CH$_2$CF$_2$SO$_3$ and 
$\kappa_{\textrm{L}}$-(ET)$_2$Ag(CF$_3$)$_4$(1-bromo-1,2,-dichloroethane), 
the similar increase of $T_{\textrm{c}}$ in the corresponding deuterated 
salts has been observed.\cite{918}
Thus deuterium substitution in the ET-based SC salts will not give the normal 
BCS isotope effect in general.\cite{isotope}

Taking account of the chemical pressure effect described above and 
the negative slope of $T_{\textrm{c}}$ with pressure in 
the electronic phase diagram of the $\kappa$-(ET)$_2X$ system,\cite{232,320,sasaki}
$T_{\textrm{c}}$ is expected to increase by deuterium substitution 
in the Cu[N(CN)$_2$]Br salt.
However, a \textit{decrease} of $T_{\textrm{c}}$ has been 
found.\cite{247,138}
This contradiction in the variation of $T_{\textrm{c}}$ is unclear, while it may 
be related to the non-bulk SC property in the deuterated salt.

To control chemical pressure, partial deuteration of ET
molecules has been studied in the Cu[N(CN)$_2$]Br salt.\cite{217,203,790,818}
This method provides a systematic variation based on the uniform donor molecules,
but only three kinds of the partially-deuterated ET have been synthesized 
between the d8- and h8-ET molecules.
Moreover the preparation of such unique donors may not always be 
convenient for making crystalline samples large enough for  
physical measurements.

As another method to apply chemical pressure to the Cu[N(CN)$_2$]Br salt,
we here investigate the d8-ET molecular substitution for 
the h8-ET molecule: $\kappa$-[(h8-ET)$_{1-x}$(d8-ET)$_{x}$]$_2$Cu[N(CN)$_2$]Br.
This simplified method is capable of giving systematic and wide range ($x$ = 0 -- 1) 
substitution and controlling chemical pressure finely near the Mott boundary.
After cooling slowly, the increase of $T_{\textrm{c}}$ by about 0.1 K is 
observed up to $x \sim$ 0.1, which is evidently caused by the chemical 
pressure effect.
Neither reduction of $T_{\textrm{c}}$ nor suppression of 
the SC volume fraction is found below $x \sim$ 0.5.
This indicates that the effect of disorder by substitution is 
negligible in the present system.

The single crystals of 
$\kappa$-[(h8-ET)$_{1-x}$(d8-ET)$_{x}$]$_2$Cu[N(CN)$_2$]Br were grown by 
a standard electrocrystallization method. 
The crystals with nominal concentrations of $x$ = 0, 0.01, 0.05, 0.1, 
0.2, 0.25, 0.3, 0.4, 0.5, 0.6, 0.7, 0.75, 0.8, 0.9, and 1 were prepared.
In this paper the concentration $x$ is used as the nominal value, because 
the actual value $x$ of each crystal has not been investigated experimentally.
Typical dimensions of the samples obtained were $\sim 1\times1\times0.2$ 
mm$^3$ ($\sim$ 0.4 mg).
For each single crystal, static magnetization measurements were 
performed with  a SQUID magnetometer (Quantum Design, MPMS-5).
The magnetic-field of 3 Oe was applied perpendicular to the conduction 
plane.
The data were taken with warming up after zero-field-cooling from above 
$T_{\textrm{c}}$ (ZFC: shielding curve) and then with cooling under the 
magnetic-field (FC: Meissner curve). 
The samples were cooled with a rate of $\sim$100 K/min from room 
temperature to 15 K, giving ``quenched'' state.
After a series of the measurements, the systems were warmed up to 100 K,
and the other cooling of 0.2 K/min (slow-cooled) was then carried out.

\begin{figure}
\begin{center}
\includegraphics[width=6cm]{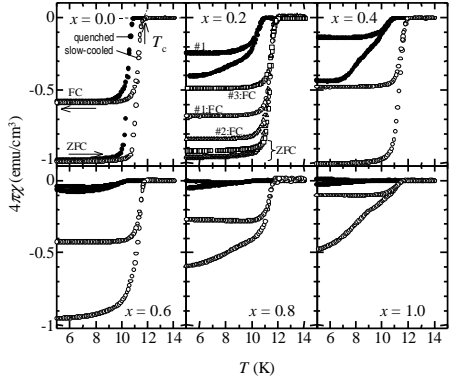}
 \caption{\label{Fig:kaiT}
 Temperature dependence of the magnetic susceptibility in 
$\kappa$-[(h8-ET)$_{1-x}$(d8-ET)$_{x}$]$_2$Cu[N(CN)$_2$]Br, where $x$ = 0, 
 0.2, 0.4, 0.6, 0.8, and 1. 
 Open and filled circles are the slow-cooled and quenched data, respectively.
 In the panel of $x$ = 0.2, the slow-cooled data in three samples are 
 displayed (\#1, 2, and \#3).
 }
\end{center}
\end{figure}

%Results and discussion:
%(1) kai-T
Figure \ref{Fig:kaiT} shows the temperature dependence of the static 
susceptibility (4$\pi\chi$) for selected several concentrations.
The demagnetization factor was corrected using an ellipsoidal approximation.
The SC transition temperatures were defined as the intercept of the 
extrapolated lines of the normal and SC states.
The results for $x$ = 0 and 1, corresponding to the pure h8- and d8-ET 
salts, respectively, are consistent with the literatures.\cite{203,sasaki,1030}
In $x$ = 0, $T_{\textrm{c}}$ is 11.83 $\pm$ 0.03 K (slow-cooled) and 
10.93 $\pm$ 0.05 K (quenched) with a sharp transition width.
The shielding curves indicate almost the full-volume fraction 
of superconductivity for both cooling-rates.
On the other hand, the broad SC transition emerges in $x$ = 1 with 
$T_{\textrm{c}}$ = 11.61 $\pm$ 0.06 K (slow-cooled) and 9.94 $\pm$ 0.15 K (quenched).
The incomplete shielding volume even in the slow-cooled state reflects 
the coexistence with a non-SC phase (AFI state\cite{144,749}).
The SC state is further suppressed by quenching.
In the panel of $x$ = 0.2, the slow-cooled data in three samples are 
displayed (\#1, 2, and \#3; taken from the same batch).
The ZFC curve and $T_{\textrm{c}}$ are well reproduced, while
the FC curves show the sample-dependent behavior.
The similar variation appears even in the pure h8-ET salt (not shown in 
Figure). 
Thus it is not the problem on the substitution, but 
may reflect the inevitable sample-dependence of the vortex-pinning 
force\cite{1030} in the low magnetic-field region, such as surface 
pinning.
In the following, we focus on the ZFC data of the slow-cooled state in 
$x$ = 0.4 -- 0.8.
The ZFC curve in $x$ = 0.4 shows the full-volume fraction within the 
accuracy of the measurements.
Almost the same sharpness of the transition width with $x$ = 0
can provide evidence of the complete SC state without any non-SC phase.
In contrast to the data in $x$ = 0.4, the transition width becomes
slightly broad in $x$ = 0.6 and meanwhile the temperature dependent 
behavior of the ZFC curve remains even at 5.0 K.
In $x$ = 0.8 the SC volume fraction is rather suppressed 
and $T_{\textrm{c}}$ decreases as well as the case in $x$ = 1.

\begin{figure}
\begin{center}
\includegraphics[width=6cm]{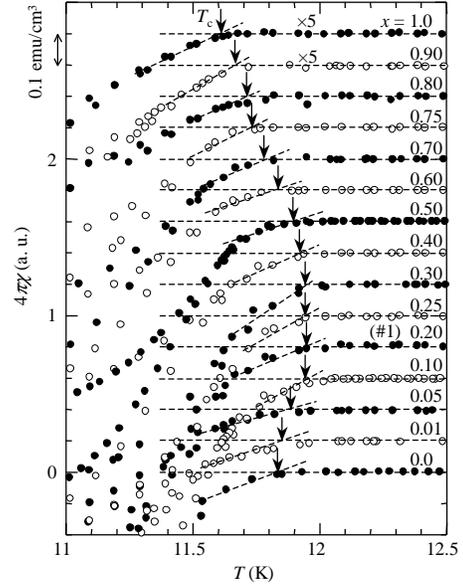}
 \caption{\label{Fig:Tcx}
 Temperature dependence of the magnetic susceptibility in 
$\kappa$-[(h8-ET)$_{1-x}$(d8-ET)$_{x}$]$_2$Cu[N(CN)$_2$]Br around 
$T_{\textrm{c}}$ in the slow-cooled state.
Arrows indicate $T_{\textrm{c}}$, defined as the intercept of the 
extrapolated broken lines of the normal and SC states.
 }
\end{center}
 \end{figure}

\begin{figure}
\begin{center}
\includegraphics[width=6cm]{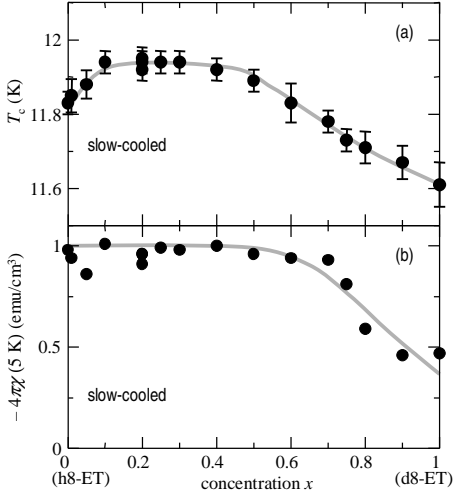}
 \caption{\label{Fig:slow}
 (a) Superconducting transition temperature and (b) values of 
 $-4\pi\chi$ at 5.0 K as a function of concentration 
 $x$ in the slow-cooled state of 
 $\kappa$-[(h8-ET)$_{1-x}$(d8-ET)$_{x}$]$_2$Cu[N(CN)$_2$]Br. 
 The bold curves are guides to the eye.
 }
\end{center}
 \end{figure}

\begin{figure}
\begin{center}
\includegraphics[width=6cm]{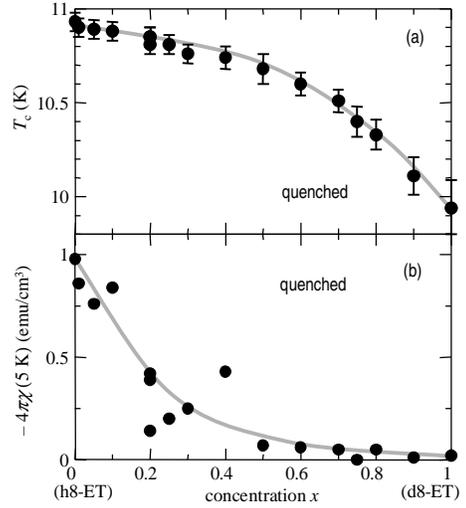}
 \caption{\label{Fig:quench}
 (a) Superconducting transition temperature and (b) values of 
 $-4\pi\chi$ at 5.0 K as a function of concentration 
 $x$ in the quenched state of 
 $\kappa$-[(h8-ET)$_{1-x}$(d8-ET)$_{x}$]$_2$Cu[N(CN)$_2$]Br.  
 The bold curves are guides to the eye.
 }
\end{center}
\end{figure}

%(2)Tc(x)  and Volume fraction for slow-cooled
Figure \ref{Fig:Tcx} shows the slow-cooled data of 4$\pi\chi(T)$ around 
$T_{\textrm{c}}$.
As indicated by arrows, $T_{\textrm{c}}$ obviously increases from $x$ = 0 to
$\sim$ 0.1.
Between $x$ = 0.2 and 0.4, $T_{\textrm{c}}$ reaches a maximum value,
and then gradually starts to decrease above $x \sim$ 0.5 toward the 
value of the pure d8-ET salt.
To exhibit the convex curvature of $T_{\textrm{c}}$ as a function of $x$,
the $x$-dependence of $T_{\textrm{c}}$ in the slow-cooled state is
shown in Fig. \ref{Fig:slow}(a).
In addition, the values of $-4\pi\chi$ at 5.0 K as the SC volume 
fraction is displayed in Fig. \ref{Fig:slow}(b). 
One can see the bulk SC state in $x \leq$ 0.5.
Above $x \sim$ 0.5, both $T_{\textrm{c}}$ and volume fraction seem to 
decrease, reflecting the existence of non-SC phases, 
although the large ambiguity in the demagnetization factor to
estimate the SC volume fraction obscures the maximum $x$ of
the bulk SC state.
On the other hand, in the quenched state, both $T_{\textrm{c}}$ and SC 
volume fraction, shown in Figs. \ref{Fig:quench}(a) and (b),
monotonically decrease with $x$,  
in which the SC volume fraction is steeply suppressed in the small $x$ 
region.

In the present deuterium substitution,
the characteristic increase of $T_{\textrm{c}}$ observed up to $x \sim$ 
0.1 in the slow-cooled state is evidently caused by the chemical 
pressure effect, as well as the case of the $\kappa$-(d8-ET)$_2$Cu(NCS)$_2$.
Note that the perfect SC volume fraction of the slow-cooled state 
appears in these concentrations [Fig. \ref{Fig:slow}(a)]. 
Conversely, $T_{\textrm{c}}$ can be reduced by the following mechanisms: 
(1) the normal (BCS) isotope effect, (2) a disorder effect introduced by 
molecular substitution, and (3) a magnetic impurity effect coming from 
the coexisting non-SC phase, which will be in the AFI state.
We propose that the third one is attributed to the origin on
the decrease of $T_{\textrm{c}}$ with $x$ in the present system.
As shown in Figs. \ref{Fig:slow} and \ref{Fig:quench}, the decrease
of $T_{\textrm{c}}$ is accompanied with the suppression of 
the SC volume fraction in both cooling conditions.
This indicates that the non-SC phase develops with increasing $x$,
leading to the reduction of $T_{\textrm{c}}$.
An assumption that these non-SC phases are in the AFI states 
will be reasonable, as it is proved in the d8-ET salt ($x$ = 1).\cite{144,749}
Since magnetic impurities reduce $T_{\textrm{c}}$ in singlet-pairing 
superconductors,\cite{977} $T_{\textrm{c}}$ therefore decreases with developing 
the non-SC phases. 
The former two mechanisms can be excluded because the first one has not 
been observed in deuterium substitution for the ET-based salts,\cite{926,918,isotope}
and thus the present system will be also not the case. 
Although the second one is widely seen in anion substitution,
such as $\beta$-(h8-ET)$_2$(I$_3$)$_{1-x}$(IBr$_2$)$_x$\cite{IBr} and 
$\kappa$-(h8-ET)$_2$Cu[N(CN)$_2$]Br$_{1-x}$I$_{x}$,\cite{1029}
this will be also ruled out because of the less difference between H and 
D atoms than that between the anions. 
This point is discussed in the following, taking account of the 
partially-deuterated system.

%(4)no disorder effect
The effect on the superconductivity by disorder in the present 
substitution is negligible, whereas they may exist on a scale of the 
molecular size. 
In the partially-deuterated ET system,\cite{203}
after slow cooling, $T_{\textrm{c}}$ starts to decrease between d-[2,2] 
and d-[3,3] salts with increasing degree of deuteration, which contain 
50\% and 75\% deuterium atoms in a ET molecule, respectively.
In both the d-[2,2] and d-[3,3] salts, almost the full-volume fraction 
after slow cooling is observed, while the bulk superconductivity 
is suppressed by fast cooling.
These trends are very similar to the present substitution system.
Thus our experimental data guarantee the minimal effect by disorder, 
which is almost comparable to the partially-deuterated system. 

\begin{figure}[h]
\begin{center}
\includegraphics[width=6cm]{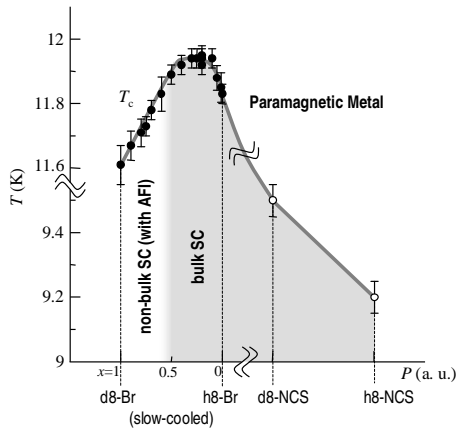}
 \caption{\label{Fig:PD}
 Schematic phase diagram in $\kappa$-[(h8-ET)$_{1-x}$(d8-ET)$_{x}$]$_2$Cu[N(CN)$_2$]Br.
The horizontal axis indicates the degree of chemical pressure.
The same scale unit is used in both the salts.
The present results are plotted between the h8-ET ($x$ = 0) and d8-ET 
salts ($x$ = 1).
The data of $\kappa$-(ET)$_2$Cu(NCS)$_2$ are taken from ref. \citen{926}.
The bold line is a guide to the eye. 
 }
\end{center}
 \end{figure}

A schematic phase diagram on the present deuterium substitution
system is depicted in Fig. \ref{Fig:PD}.
The horizontal axis indicates the degree of chemical pressure.
To compare with the present results, the data of the Cu(NCS)$_2$ salt
taken from ref. \citen{918} are also plotted.
The same scale unit is used in both Cu[N(CN)$_2$]Br and Cu(NCS)$_2$ salts, 
in which the present data are plotted between the h8-ET ($x$ = 0) and d8-ET 
salts ($x$ = 1).
The most feature in this phase diagram is the boundary 
between the bulk SC and non-bulk SC (with AFI) phases at around $x$ = 0.5.
This boundary is expected to be the Mott transition induced by 
chemical pressure, in which  both $T_{\textrm{c}}$ and SC volume 
fraction start to decrease.
In the Cu(NCS)$_2$ salt, $T_{\textrm{c}}$'s are 9.20 $\pm$ 0.05 K and 
9.50 $\pm$ 0.05 K for the hydrogenated and deuterated salts, respectively.\cite{918}
Thus deuterium substitution in this salt increases $T_{\textrm{c}}$ by 
0.3 $\pm$ 0.1 K.
The degree of chemical pressure giving such an increase of $T_{\textrm{c}}$ is 
estimated to be about 40 bar as uniaxial-pressure
toward inter-plane direction.\cite{926}
We note that the variation of $T_{\textrm{c}}$ in the Cu[N(CN)$_2$]Br 
salt is on the same order with the Cu(NCS)$_2$ salt: 
the increase by about 0.1 K up to $x \sim$ 0.1 and decrease by about
0.3 K from $x$ = 0.5 to 1.
This demonstrates that deuterium substitution in the $\kappa$-ET 
system gives the same degree of chemical pressure, as suggested by 
Schlueter et al.\cite{918}
Moreover, further decrease of $T_{\textrm{c}}$ is observed in the 
quenched state: about 1 K from $x$ = 0 to 1 [see Fig. \ref{Fig:quench}(a)]. 
This can be interpreted as the influence of the much more non-SC phase 
as shown in Fig. \ref{Fig:quench}(b), giving rise to further
magnetic-impurity scattering than that in the slow-cooled state.

%Summary
In summary, we have investigated the superconductivity in the d8-ET molecular
substitution system by means of the magnetic susceptibility measurements.
The measurements were performed in the two cooling conditions of the slow-cooled 
(0.2 K/min) and quenched ($\sim$ 100 K) states.
On the one hand, in the slow-cooled state, the chemical pressure effect 
leads to an increase of $T_{\textrm{c}}$ by about 0.1 K.
Neither reduction of $T_{\textrm{c}}$ nor suppression of the 
SC volume fraction is observed below $x \sim$ 0.5.
This demonstrates that the effect of disorder by substitution is 
sufficiently negligible in the present system.
With further increase of $x$, both $T_{\textrm{c}}$ and 
SC volume fraction start to decrease toward the values in $x$ =1.
On the other hand, the data of the quenched state between $x$ = 0 and 1 
qualitatively resemble the behavior of the slow-cooled state in $x \geq$ 0.5; 
both $T_{\textrm{c}}$ and SC volume fraction monotonically decrease, 
while the degrees of the $T_{\textrm{c}}$'s reduction or the SC volume 
suppression are much larger than that in  
the slow-cooled state. 
The reduction of $T_{\textrm{c}}$ is interpreted as a magnetic impurity 
effect coming from the coexisting non-SC (AFI) phase.
As a result, this simple substitution method makes it possible to 
control chemical pressure with minimal disorder.
It is a useful technique to approach the Mott boundary at 
ambient pressure because the samples of any concentration $x$ can be 
easily prepared.

\section*{Acknowledgment}
This research was partly supported by the Ministry of Education,
Science, Sports and Culture, Grant-in-Aid for Encouragement of Young
Scientists (NY, Grant No. 14740373), and for Scientific Research (C) (TS, 
Grant No. 15540329).


\begin{thebibliography}{99}
%%%%%% References %%%%%%%%%%%%%
\bibitem{Ishiguro}T. Ishiguro, K. Yamaji and G. Saito:
Organic Superconductors, 2nd ed. (Springer, Berlin, 1998).

\bibitem{235}H. Ito, T. Ishiguro, M. Kubota and G. Saito:
J. Phys. Soc. Jpn. \textbf{65} (1996) 2987.

\bibitem{327}S. Lefebvre, P. Wzietek, S. Brown, C. Bourbonnais, D. J\'{e}rome, 
C. M\'{e}zi`{e}re, M. Fourmigu\'{e} and P. Batail:
Phys. Rev. Lett. \textbf{85} (2000) 5420.

\bibitem{986}D. Fournier, M. Poirier, M. Castonguay and K. D. Truong: 
Phys. Rev. Lett. \textbf{90} (2003) 127002.

\bibitem{971}P. Limelette, P. Wzietek, S. Florens, A. Georges, T. A. 
Costi, C. Pasquier, D. J\'{e}rome, C. M\'{e}zi`{e}re and P. Batail:
Phys. Rev. Lett. \textbf{91} (2003) 016401.

\bibitem{346}H. Kino and H. Fukuyama:
J. Phys. Soc. Jpn. \textbf{65} (1996) 2158.

\bibitem{144}A. Kawamoto, K. Miyagawa and K. Kanoda:
Phys. Rev. B \textbf{55} (1997) 14140.

\bibitem{749}K. Miyagawa, A. Kawamoto and K. Kanoda: 
Phys. Rev. Lett. \textbf{89} (2002) 017003.

\bibitem{926}A. M. Kini, K. D. Carlson, H. H. Wang, J. A. Schlueter, J. 
D. Dudek, S. A. Sirchio, U. Geiser, K. R. Lykke and J. M. Williams: 
Physica C \textbf{264} (1996) 81.

\bibitem{918}J. A. Schlueter, A. M. Kini, B. H. Ward, U. Geiser, H. H. 
Wang, J. Mohtasham, R. W. Winter and G. L. Gard:
Physica C \textbf{351} (2001) 261.

\bibitem{isotope}Other isotopic substitutions ($^{13}$C and $^{34}$S) 
in $\kappa$-(BEDT-TTF)$_2$Cu(NCS)$_2$ has been investigated, resulting 
in the normal isotope effect on $T_{\textrm{c}}$.

\bibitem{232}K. Kanoda: 
Physica C \textbf{282-287} (1997) 299.

\bibitem{320}R. H. Mckenzie:
Science \textbf{278} (1997) 820.

\bibitem{sasaki}T. Sasaki, N. Yoneyama, A. Matsuyama and N. Kobayashi: 
Phys. Rev. B \textbf{65} (2002) 060505.

\bibitem{247}M. Tokumoto, N. Kinoshita, Y. Tanaka and H. Anzai:
J. Phys. Soc. Jpn. \textbf{60} (1991) 1426.

\bibitem{138}H. Ito, M. Watanabe, Y. Nogami, T. Ishiguro, T. Komatsu, G. 
Saito and N. Hosoito:
J. Phys. Soc. Jpn. \textbf{60} (1991) 3230.

\bibitem{217}A. Kawamoto, H. Taniguchi and K. Kanoda:
J. Am. Chem. Soc. \textbf{120} (1998) 10984.

\bibitem{203}H. Taniguchi, A. Kawamoto and K. Kanoda:
Phys. Rev. B \textbf{59} (1999) 8424.

\bibitem{790}Y. Nakazawa, H. Taniguchi, A. Kawamoto and K. Kanoda:
Phys. Rev. B \textbf{61} (2000) R16295.

\bibitem{818}H. Taniguchi, K. Kanoda and A. Kawamoto:
Phys. Rev. B \textbf{67} (2003) 014510.

\bibitem{1030}N. Yoneyama, T. Sasaki, T. Nishizaki and N. Kobayashi:
J. Phys. Soc. Jpn. \textbf{73} (2004) 184.

\bibitem{977} B. J. Powell and R. H. McKenzie: 
Phys. Rev. B \textbf{69} 024519.

\bibitem{IBr}M. Tokumoto, H. Anzai, K. Murata, K. Kajimura and T. Ishiguro: 
J. Jpn. Appl. Phys. \textbf{26} (1987) 1977.

\bibitem{1029}N. D. Kushch, L. I. Buravov, A. G. Khomenko, S. I. 
Pesotskii, V. N. Laukhin, E. B. Yagubskii, R. P. Shibaeva, V. E. 
Zavodnik and L. P. Rozenberg:
Synth. Metals \textbf{72} (1995) 181.

\end{thebibliography}
\end{document}